\documentclass[%
 reprint,
superscriptaddress,
 amsmath,amssymb,
 aps,
prl
]{revtex4-2}

\newcommand{\ket}[1]{\lvert#1\rangle} 

\newcommand{\da}{\downarrow}

\newcommand{\ketda}{\ket{\downarrow}}
\newcommand{\ketua}{\ket{\uparrow}}

\usepackage{graphicx}
\usepackage{dcolumn}
\usepackage{bm}
\usepackage{hyperref}
\usepackage[mathlines]{lineno}

\begin{document}

\title{Demonstration of chronometric leveling using transportable optical clocks beyond laser coherence limit}

\author{Yi Yuan}
\affiliation{State Key Laboratory of Magnetic Resonance and Atomic and Molecular Physics, Innovation Academy for Precision Measurement Science and Technology, Chinese Academy of Sciences, Wuhan 430071, China}
\affiliation{Key Laboratory of Atomic Frequency Standards, Innovation Academy for Precision Measurement Science and Technology, Chinese Academy of Sciences, Wuhan 430071, China}
\affiliation{University of the Chinese Academy of Sciences, Beijing 100049, China}
\author{Kaifeng Cui}
\affiliation{State Key Laboratory of Magnetic Resonance and Atomic and Molecular Physics, Innovation Academy for Precision Measurement Science and Technology, Chinese Academy of Sciences, Wuhan 430071, China}
\affiliation{Key Laboratory of Atomic Frequency Standards, Innovation Academy for Precision Measurement Science and Technology, Chinese Academy of Sciences, Wuhan 430071, China}
\author{Daoxin Liu}
\affiliation{State Key Laboratory of Magnetic Resonance and Atomic and Molecular Physics, Innovation Academy for Precision Measurement Science and Technology, Chinese Academy of Sciences, Wuhan 430071, China}
\affiliation{Key Laboratory of Atomic Frequency Standards, Innovation Academy for Precision Measurement Science and Technology, Chinese Academy of Sciences, Wuhan 430071, China}
\author{Jinbo Yuan}
\affiliation{State Key Laboratory of Magnetic Resonance and Atomic and Molecular Physics, Innovation Academy for Precision Measurement Science and Technology, Chinese Academy of Sciences, Wuhan 430071, China}
\affiliation{Key Laboratory of Atomic Frequency Standards, Innovation Academy for Precision Measurement Science and Technology, Chinese Academy of Sciences, Wuhan 430071, China}
\author{Jian Cao}
\email{caojian@apm.ac.cn}
\affiliation{State Key Laboratory of Magnetic Resonance and Atomic and Molecular Physics, Innovation Academy for Precision Measurement Science and Technology, Chinese Academy of Sciences, Wuhan 430071, China}
\affiliation{Key Laboratory of Atomic Frequency Standards, Innovation Academy for Precision Measurement Science and Technology, Chinese Academy of Sciences, Wuhan 430071, China}
\author{Dehao Wang}
\affiliation{State Key Laboratory of Magnetic Resonance and Atomic and Molecular Physics, Innovation Academy for Precision Measurement Science and Technology, Chinese Academy of Sciences, Wuhan 430071, China}
\affiliation{Key Laboratory of Atomic Frequency Standards, Innovation Academy for Precision Measurement Science and Technology, Chinese Academy of Sciences, Wuhan 430071, China}
\affiliation{University of the Chinese Academy of Sciences, Beijing 100049, China}
\author{Sijia Chao}
\affiliation{State Key Laboratory of Magnetic Resonance and Atomic and Molecular Physics, Innovation Academy for Precision Measurement Science and Technology, Chinese Academy of Sciences, Wuhan 430071, China}
\affiliation{Key Laboratory of Atomic Frequency Standards, Innovation Academy for Precision Measurement Science and Technology, Chinese Academy of Sciences, Wuhan 430071, China}
\author{Hualin Shu}
\affiliation{State Key Laboratory of Magnetic Resonance and Atomic and Molecular Physics, Innovation Academy for Precision Measurement Science and Technology, Chinese Academy of Sciences, Wuhan 430071, China}
\affiliation{Key Laboratory of Atomic Frequency Standards, Innovation Academy for Precision Measurement Science and Technology, Chinese Academy of Sciences, Wuhan 430071, China}
\author{Xueren Huang}
\email{hxueren@apm.ac.cn}
\affiliation{State Key Laboratory of Magnetic Resonance and Atomic and Molecular Physics, Innovation Academy for Precision Measurement Science and Technology, Chinese Academy of Sciences, Wuhan 430071, China}
\affiliation{Key Laboratory of Atomic Frequency Standards, Innovation Academy for Precision Measurement Science and Technology, Chinese Academy of Sciences, Wuhan 430071, China}
\affiliation{Wuhan Institute of Quantum Technology, Wuhan 430206, China}

\date{\today}

\begin{abstract}
Optical clock network requires the establishment of optical frequency transmission link between multiple optical clocks, utilizing narrow linewidth lasers. Despite achieving link noise levels of 10${^{-20}}$, the final accuracy is limited by the phase noise of the clock laser. Correlation spectroscopy is developed to transmit frequency information between two optical clocks directly, enabling optical clock comparison beyond the phase noise limit of clock lasers, and significantly enhancing the measurement accuracy or shorten the measurement time. In this letter, two compact transportable ${^{40}}$Ca${^+}$ clocks are employed to accomplish the correlation spectroscopy comparison, demonstrating an 10 cm level measurement accuracy of chronometric leveling using a mediocre clock laser with linewidth of 200 Hz. The relative frequency instability reaches $6.0\times10{^{-15}}/\sqrt{\tau/s}$, which is about 20 times better than the result with Rabi spectroscopy using the same clock laser. This research greatly reduces the harsh requirements on the performance of the clock laser, so that an ordinary stable-laser can also be employed in the construction of optical clock network, which is essential for the field applications, especially for the chronometric leveling.
\end{abstract}

\maketitle


\section{\label{sec:level1}Introduction}
Optical atomic clocks with fractional systematic uncertainties at the 10$^{-18}$ level or below\cite{PhysRevLett.104.070802,ushijima2015cryogenic,nicholson2015systematic,PhysRevLett.116.063001,mcgrew2018atomic,takamoto2020test,PhysRevApplied.17.034041,zhiqiang2023176lu+} are the most precise measurement devices of any kind, driven greatly by the progress in ultrastable lasers\cite{PhysRevLett.123.173201}. In metrology, a single clock can serve as the absolute frequency standard in principle\cite{Gill_2005,Falke_2011}. However, numerous other applications require two or more optical clocks to establish a network, facilitating the desired observation by comparing frequency difference between each clock\cite{lisdat2016clock,riehle2017optical,boulder2021frequency,derevianko2014hunting,PhysRevD.94.124043}. In such comparison-mode scenarios, the frequency difference measurement stability is limited by the laser noise, which restricts the clock probe time and introduces Dick effect noise\cite{quessada2003dick}. Better clock laser system, e.g. locking to cryogenic silicon cavity has been built to extend the clock probe time and increase the duty cycle\cite{oelker2019demonstration}. However, these kinds of high-performance ultrastable cavities are always bulky, and sensitive to the environment, so the robustness is inadequate to satisfy a wide range of applications such as chronometric leveling in the field, even in harsh environments for global geodetic observation networks\cite{takamoto2020test,takano2016geopotential,grotti2018geodesy,mehlstaubler2018atomic,lion2017determination}. Furthermore, the narrow linewidth laser serves solely as the carrier of the frequency difference information between clock transitions, and not as the key information required in the optical clock comparisons. Consequently, it is possible to achieve clock comparison far beyond the laser coherence time limit by appropriate methods. 

To achieve frequency comparison beyond clock laser's coherent time, it is necessary to measure the evolution of coherence between the optical clocks. One of these methods is to prepare an entangled state of two trapped ions through fiber network, which requires high-fidelity entanglement preparation between remote trapped ion systems\cite{PhysRevLett.123.203001,colombo2022entanglement,nichol2022elementary}. In contrast, correlation spectroscopy is simple and robust for building a global optical clock network as it takes up fewer resources\cite{chwalla2007precision,PhysRevLett.125.243602}. With this method, coherence between two independent ${^{27}}$Al${^+}$ clocks at probe duration as long as 8 s on the clock transition was observed\cite{PhysRevLett.125.243602}. According to the contrast in the correlation spectrum, an estimated instability of $1.8\times10^{-16}/\sqrt{\tau/s}$ is possible in the ion clocks comparison and thus comparable to the performance of using optical lattice clocks\cite{PhysRevLett.125.243602}.

In this research, we demonstrated the closed-loop operation of clocks based on correlation spectroscopy  method and applied it to the chronometric leveling using a clock laser with poor coherence. Using two transportable ${^{40}}$Ca${^+}$ optical clocks, a 729 nm clock laser with an intentionally destroyed coherence time of 2.6 ms, corresponding to a linewidth of 200 Hz is employed to achieve 100 ms Ramsey probe durations for the clocks comparison. The chronometric leveling results are consistent  with the values obtained by the traditional method of Rabi spectroscopy, and the accuracy is limited to the level of 10 cm due to the $1\times10^{-17}$ systematic uncertainties of both clocks\cite{liu2023laboratory,YuanjinboAPL}, the instability of frequency difference measurement has been significantly improved to $6.0\times10{^{-15}}/\sqrt{\tau/s}$, which is about 20 times better than the result obtained with clocks operated at 2 ms Rabi probe time using the same intentionally destroyed clock laser.
\section{Experiment methods and results}
For simplicity, only two sets of single-ion optical clock are involved here. Assuming that the two ions have the same ground state $\ketda$ and excited state $\ketua$, a Ramsey-type clock comparison scheme is described as the following three steps: (1) Prepare a quantum superposition state $\ket{\phi} = (\ketda+\ketua)/\sqrt{2}$ of each ion with a $\pi/2$ clock pulse; (2) Turn off the clock laser and let the ions distributed in different traps to evolve independently for a preset time of $\tau_R$; (3) Apply the second $\pi/2$ pulse to the ion and measure the population $P_\downarrow$ of each ion in the $\ketda$ state. If there is a mismatch between the frequency $\nu_L$ of the clock laser and the atomic resonant frequency $\nu_0$, and $\delta = \nu_L-\nu_0$, the ion will accumulate a phase $\varphi = 2\pi \tau_R \delta$ during the Ramsey probe duration. Therefore, the interference fringe $P_\da = A \cos(2\pi\delta\tau_R + \phi_0)$ will be observed, where A is defined as the fringe contrast and $\phi_0$ is the initial phase associated with laser detuning.
When both ions are interrogated simultaneously using the common laser, we can measure the parity $\hat{\Pi}$ instead of the transition probability to realize a correlation spectroscopy. Here $\hat{\Pi}$ is defined as the time-average of the parity operator $\hat{\Pi}=\hat{\sigma}_{z,1} \hat{\sigma}_{z,2}$, where $\hat{\sigma}_{z,i}$ represent the Pauli operator of trapped ion in the i-th trap. The measurement result is the product of the oscillations of the two systems\cite{PhysRevLett.125.243602}:
\begin{equation}
    \langle \hat{\Pi} \rangle = 
        \frac{1}{2} [
        \cos(2\pi\delta_+\tau_R + \phi_+) + \cos(2\pi\delta_-\tau_R + \phi_-)
\end{equation}
where $\delta_{\pm} = \delta_1 \pm \delta_2$ and $\phi_{\pm} = \phi_1 \pm \phi_2$. If the free evolution time $\tau_R$ is long enough compared to the laser coherence time, the first term in Eq. (1) will be averaged to zero. The phase decoherence of the laser is averaged due to the common mode effect of the two ion systems, leaving only the independent spontaneous emission parts of the two ion systems. Introducing a decay rate $\Gamma$ due to spontaneous emission, the correlation measurements will give the following interference fringe:
\begin{equation}
    \langle \hat{\Pi} \rangle = \frac{1}{2}e^{-\Gamma\tau_R}\cos(2\pi\delta_-\tau_R + \phi_-)
\end{equation}
The frequency difference information of the two ions appears in the interference fringe, and the common mode of the laser phase noise is suppressed. If the feedback control method is adopted to track the maximum slope points of the interference fringe, the following measurement instability can be obtained:
\begin{equation}
    \sigma(\tau) = \frac{1}{\pi\nu_0 \sqrt{\tau_R\tau}} e^{\Gamma \tau_R}    
\end{equation}
Here, the measurement instability is not limited by the coherence time of the clock laser, but only related to the spontaneous emission rate of the ion's clock transition. 
\begin{figure}
    \centering
    \includegraphics[width=0.95\linewidth]{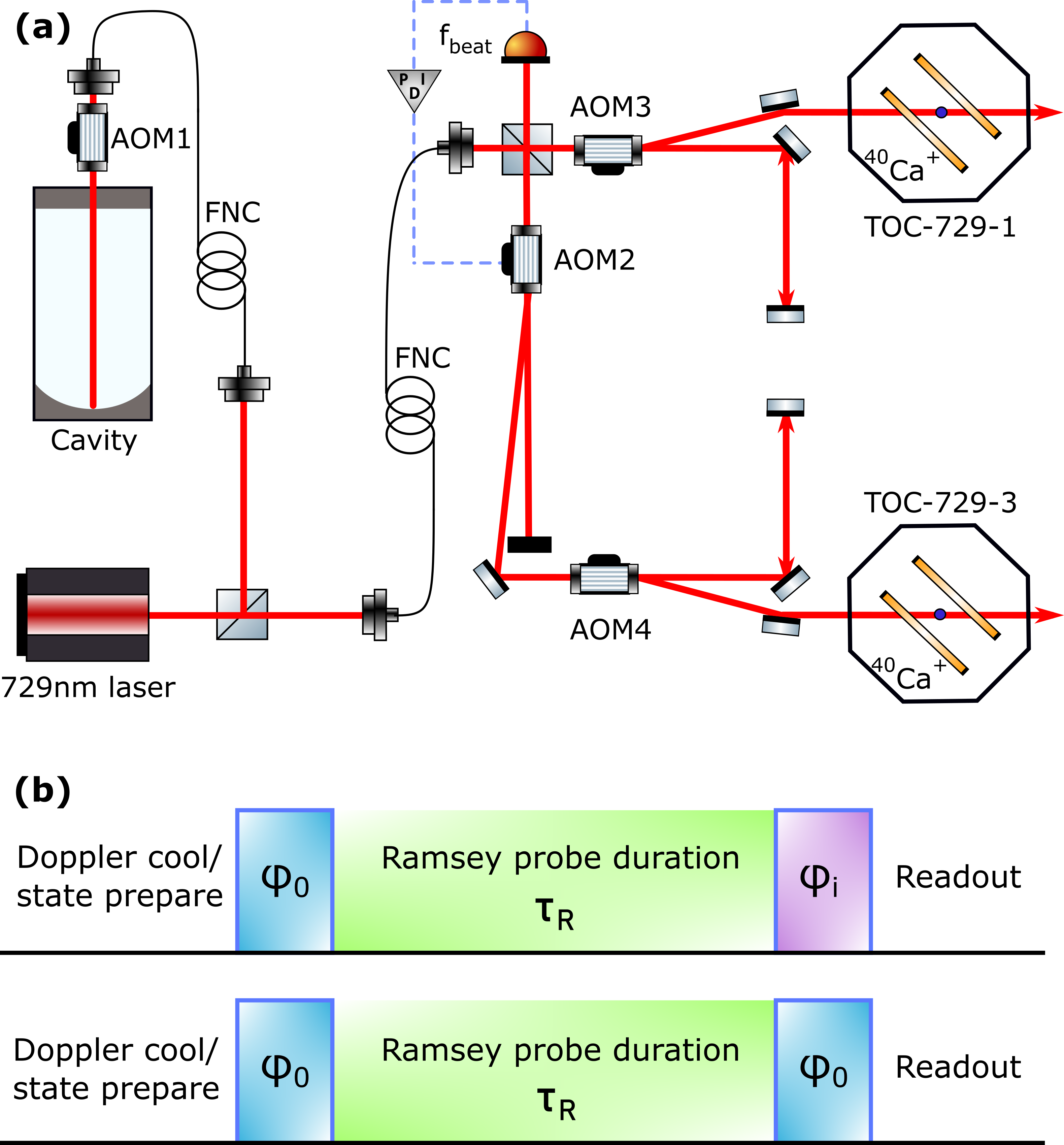}
    \caption{
    The devices and clock comparison scheme of the correlation spectroscopy experiment. (a) Two transportable optical clocks connected by a fiber noise synchronization link share the same ultrastable laser, where TOC-729-1 is placed on a fixed platform, and TOC-729-3 can move up and down. (b) Synchronous inquiry timing scheme. The ${^2}S_{1/2}\rightarrow{^2}D_{5/2}$ transitions of the two ${^{40}}$Ca${^+}$ optical clocks are excited simultaneously, but the phase of the second pulse of the two clocks is different.
    }
\end{figure}
As shown in Fig. 1, our experimental setup consists of two separate transportable optical clocks based on single trapped ${^{40}}$Ca${^+}$ ion (labeled as TOC-729-1 and TOC-729-3). The ${^2}S_{1/2}\rightarrow{^2}D_{5/2}$ clock transition of ${^{40}}$Ca${^+}$ is interrogated by a shared 729 nm laser referenced on a 10 cm ultrastable cavity. Normally, the linewidth of this clock laser is about 1 Hz, and its short-term stability reaches the level of $2\times10^{-15}$\cite{SHANG2017410}. To simulate a laser with poor coherence in this experiment, white noise modulation is added to the driver of an acousto-optical modulator (AOM 1) in the cavity path. A fiber noise cancellation system is established to synchronize the laser noise between the two trapped-ion systems. The beat signal is obtained from the zero-order optical reflections of AOM 3 and AOM 4, and its frequency noise is directed to the driver of AOM 2 by a servo to keep the coherence of the laser consistent between the two trapped ion systems. To eliminate the spatial laser noise as much as possible, the free space transmission distance of the zero-order optical reflections of each AOM is about the same as the distance from the AOM to the ion.

During the clock interrogation, the phase of the second $\pi/2$ pulse on TOC-729-1 is scanned relative to TOC-729-3, and the parity fringe is observed as shown in the inset of Fig. 2(a). Series of fringe contrasts under different free evolution time are measured, and the results are shown in Fig. 2(a), along with the laser coherence limited Ramsey spectroscopy contrast. In these measurements, due to the limited power of the 729 nm laser, the $\pi$ pulse time is approximately 400 us, so the laser coherence time is intentionally destroyed to ~2.6 ms and it's verified that the laser linewidth is degenerated to about 200 Hz by heterodyne beating with a narrow linewidth laser. There's no further destruction of the laser coherence time to a much shorter scale. But, in principle, lasers with arbitrary linewidths can be employed to achieve high-precision frequency comparison once the laser's coherence time is long enough to perform $\pi/2$ pulse but shorter than the free evolution time of the Ramsey-type interrogation.

We find that the measured coherence time using correlation spectroscopy does not reach the 1.17 s lifetime of the upper ${^2}D_{5/2}$ state of ${^{40}}$Ca${^+}$ due to the decoherence from the magnetic field noise. This can be verified by comparing the parity fringes of three Zeeman components with different magnetic sensitivities. The decoherence times of the $\Delta$m=0, -1, and -2 transitions are 797(16) ms, 317(13) ms, and 153(5) ms, respectively. Although the magnetic field noise now becomes the main limitation of the further extension of the clocks probe time, this experiment demonstrated a scheme with clock interrogation time far beyond the laser's coherent time. 
As a comparison, the coherence time of two ions in the same linear trap (TOC-729-1) is measured using  correlation spectroscopy with the $\Delta$m=0 transition (see Fig. 2 (b)). The maximal contrast is close to 0.5, which is better than the contrast obtained by two ions in different linear traps. Trapping in the same trap suppressed the ambient noise, including laser propagation noise and magnetic field noise, makes it possible to reach a coherent time of 827(252) ms.

\begin{figure}
    \centering
    \includegraphics[width=0.95\linewidth]{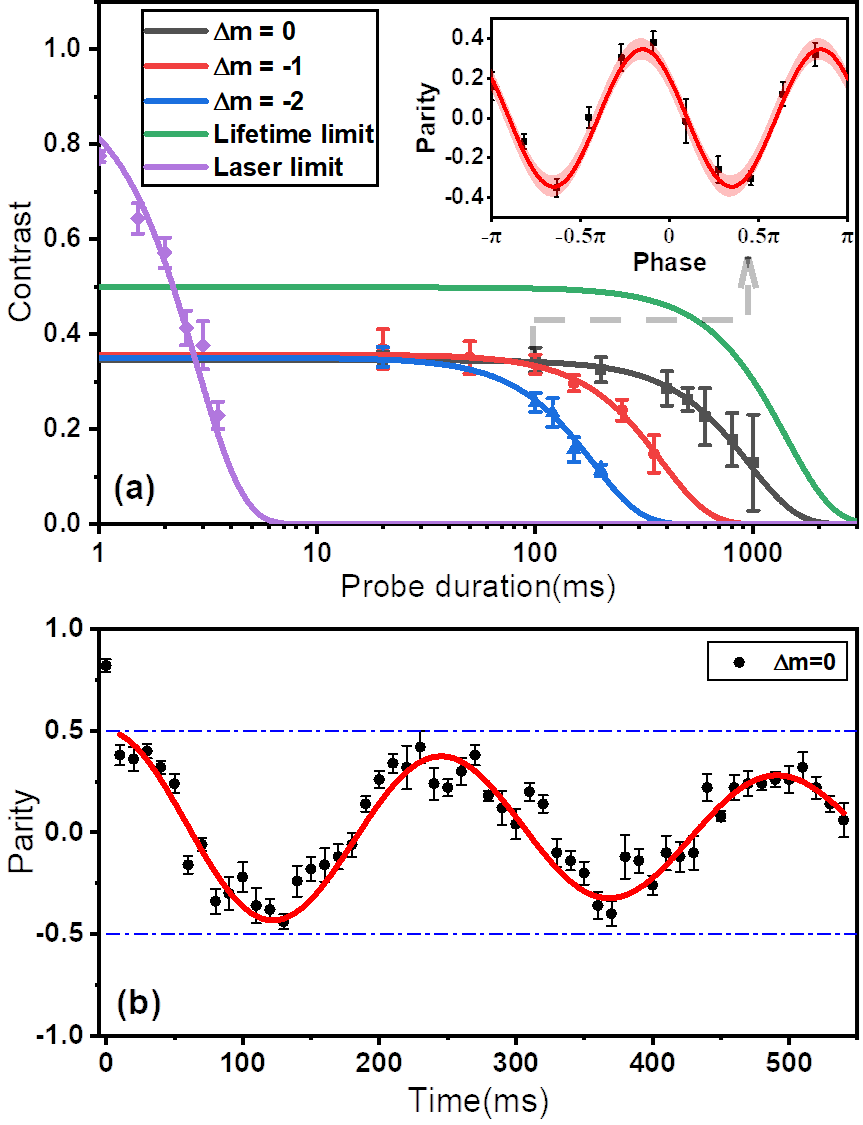}
    \caption{
    Evaluation of the two correlated optical clocks based on trapped ${^{40}}$Ca${^+}$ . (a) Contrast as a function of the probe duration. The black squares, red circles, and blue triangles are the measured contrasts of the correlation spectrum of the $\Delta$m=0, -1, -2 transitions, and the corresponding lines are the decoherence fitting curves. For comparison, the laser coherence limit measured using Ramsey spectroscopy is shown as purple diamonds, while the lifetime limit of the clock transition is shown as the green solid line. Inset: The parity fringe of the correlation spectrum at a typical 100 ms probe time of the $\Delta$m=0 transition. The horizontal axis in this inset means the phase of the second $\pi/2$ pulse of TOC-729-1, and the amplitude of fitting line represents the fringe contrast. (b) The evaluation of parity fringe as a function of probe time with two ions in the same linear trap.
    }
\end{figure}

The phase difference $\phi_-$ of the correlation spectrum is closely related to the frequency difference of the interrogated transition. So the frequency difference of the same transition of two ions can be calculated from the phase of the selected peak of the parity fringes:
\begin{equation}
    \Delta f = \frac{1}{\tau_R}({N+\frac{\phi_0}{2\pi}})
\end{equation}
where $\tau_R$ is the free evolution time, $\phi_0$ is the phase of the selected peak in the parity fringe, and $N$ is an integer constant related to the initial frequency difference. Take the results of the inset in Fig.2(a) as an example, since the magnetic field difference between TOC-729-1 and TOC-729-3 ions is about 2905(1) nT, the corresponding frequency difference of $\Delta$m=0 transition is 16302.7(11.2) Hz, the frequencies of ions' free evaluation during the Ramsey probe time are different in these two clocks. Considering that the frequency of the transitions is different between the two ions, the phase difference detected by the parity fringes under the free evolution time of 100 ms should be -1630.27(1.12)$\times2\pi$, this result is consistent with the peak phase -0.153(0.03)$\times2\pi$ of the parity fringe fitting.
\begin{figure}
    \centering
    \includegraphics[width=0.95\linewidth]{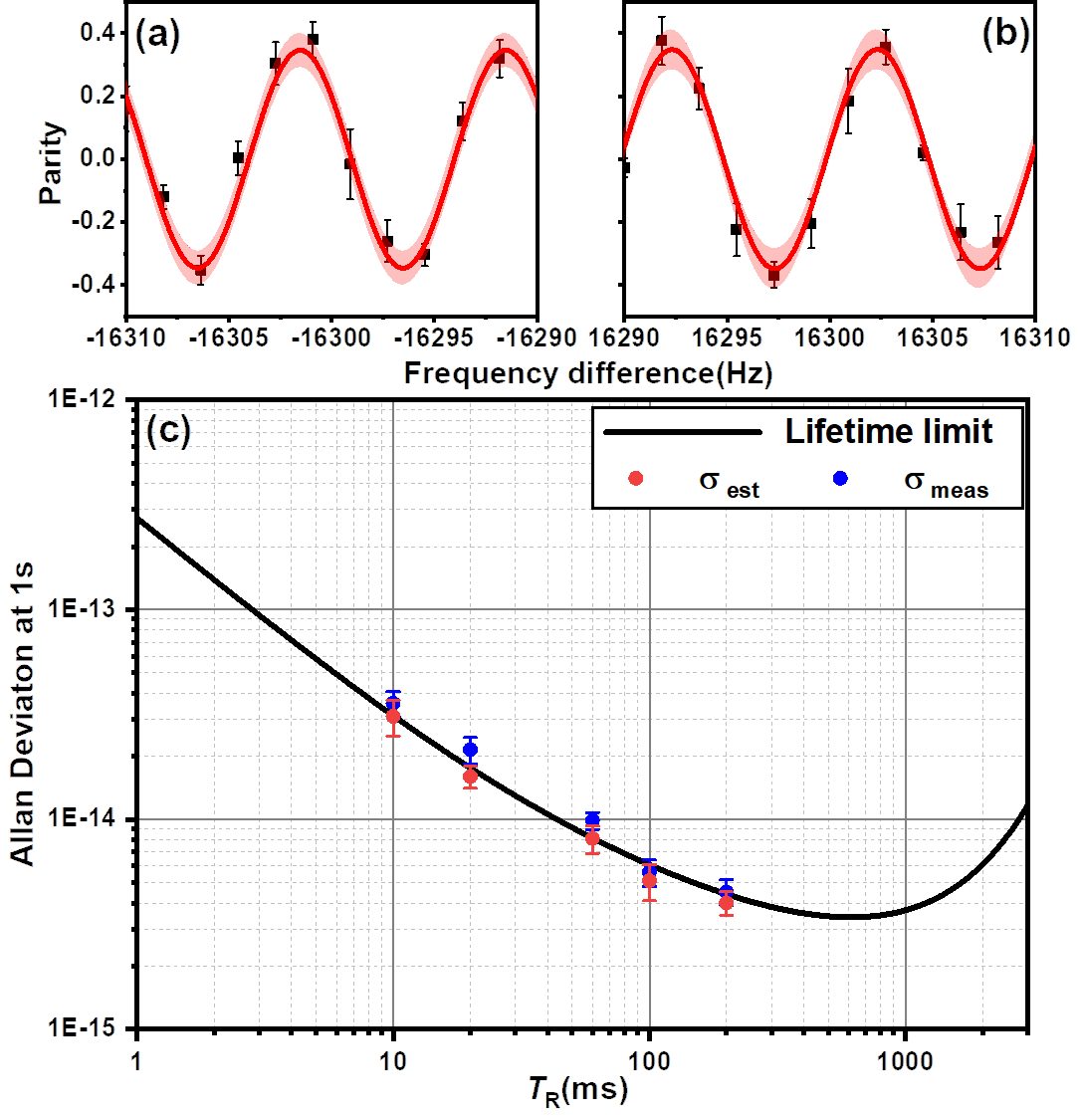}
    \caption{
    Demonstration of closed-loop operation referenced on the parity fringes. (a,b) A pair of parity fringes converted by phase scanning with $\Delta$m=0, the horizontal axis is the frequency difference of the same transition of the two ions, and the Ramsey probe time of the correlation spectrum is 100 ms. (c) Comparison of measured(blue point), estimated(red point), and theoretical(black line) instabilities of the clock comparisons using correlation spectroscopy.
    }
\end{figure}

The peak phase tracking of the parity fringes can be achieved by interrogating at the two points with a parity fringe on both sides of the specified peak. In the experiment, the inquiry of these peaks is also added to judge whether the locking is normal. Based on this, we demonstrate the closed-loop clock operation with a pair of Zeeman components.The correlation spectrum obtained from ${^2}S_{1/2}(m_J=-1/2)\rightarrow{^2}D_{5/2}(m_J=-1/2$) and ${^2}S_{1/2}(m_J=+1/2)\rightarrow{^2}D_{5/2}(m_J=+1/2)$ are shown in Fig.3(a,b) where the scanned phase has been calculated as frequency difference of each peak. Frequency comparisons with the parity fringes of $\Delta$m=0 transitions at the different interrogation time range from 10 ms to 200 ms are then demonstrated, and their self-comparison instability results are shown as $\sigma_{meas}$ in Fig.3(c). The instabilities estimated by the contrast of parity fringes and the instability calculated with the lifetime of ${^2}D_{5/2}$ are shown in Fig.3(c). The measured results is in great agreement with the estimated instability within the error range.

The frequency shift due to the electric quadrupole effect is eliminated by averaging three inner pairs of Zeeman components with magnetic sublevels $m_J$=$\pm$1/2, $\pm$3/2, and $\pm$5/2 of the upper ${^2}D_{5/2}$ state, sequentially in each cycle during the lock runs. The effective data duration of each measurement is approximately four hours, and the instability of the frequency difference data obtained by seven groups of measurements reaches 6.0$\times10{^{-15}}/\sqrt{\tau/s}$ (Fig.4(b)) It's about 20 times better than the limit measured by the traditional Rabi method under the same laser linewidth.

\begin{figure}
    \centering
    \includegraphics[width=0.95\linewidth]{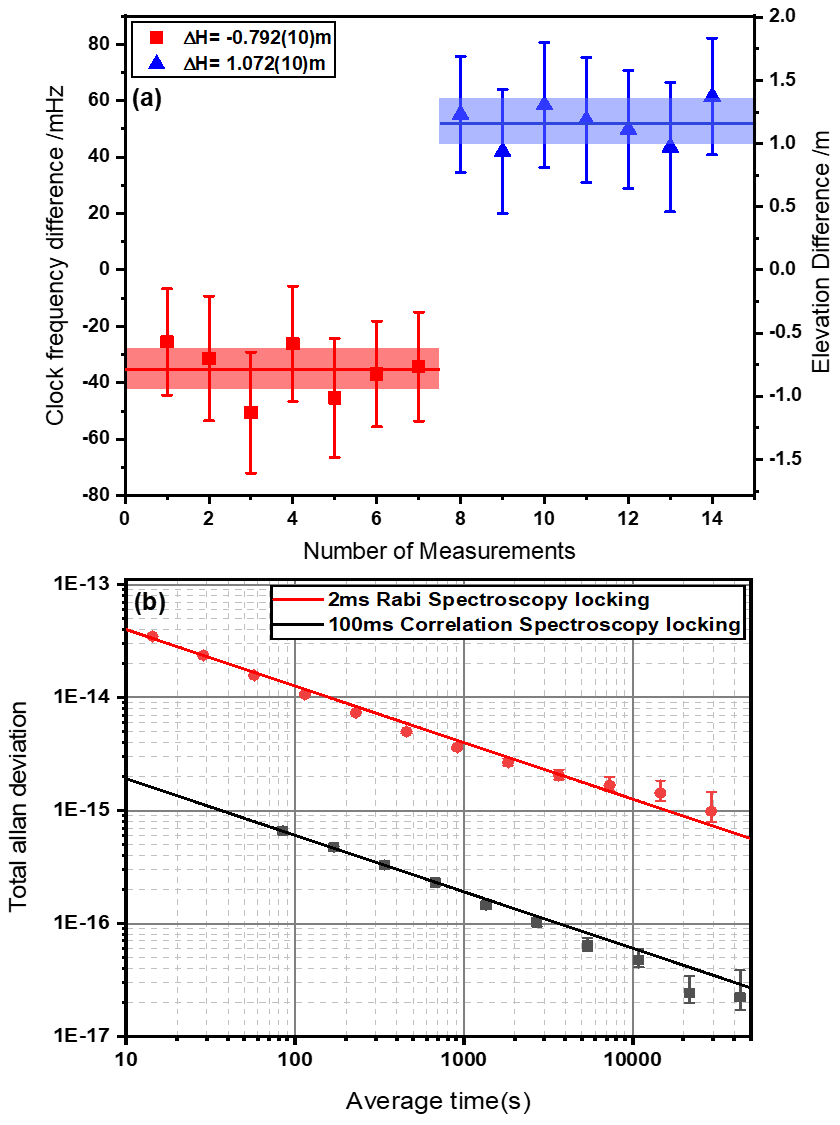}
    \caption{
    (a) Results of elevation difference measurements based on correlation spectroscopy at two different elevations. The corresponding measurement results shown by the red and blue points are corrected by the systematic shifts. The solid lines stand for the weight average of the date while the shadowed areas represent the standard deviation of the mean. (b) Relative instabilities with different frequency comparison methods. The black squares are the instability data obtained by locking on the peaks of the six parity fringes with 100 ms probe time, and the black solid line is the fitting as a result of 6.0$\times10^{-15}$/$\sqrt{\tau/s}$. The red dots are the instability of the Rabi lock with 2 ms probe time, and the red solid line is the fitting as a result of 1.2$\times10^{-13}$/$\sqrt{\tau/s}$.
    }
\end{figure}

\begin{table*}
\caption{\label{tab:table3}The elevation difference measurement results of correlation spectrum comparison under the height difference of -79.2(1.0) cm and 107.2(1.0) cm.}
\begin{ruledtabular}
\begin{tabular}{ccccc}
 $\Delta H$&Statistical frequency difference&System frequency shift&Absolute frequency difference&Elevation difference\\
 /cm&/mHz&/mHz&/mHz&/cm\\ \hline
 -79.2(1.0)&-50.5(3.3)&-15.2(6.4)&-35.3(7.2)&-78.8(16.1)\\ 
 107.2(1.0)&39.1(2.6)&-12.8(7.7)&51.9(8.2)&115.9(18.3)\\
\end{tabular}
\end{ruledtabular}
\end{table*}

After these efforts, we performed a laboratory demonstration of  chronometric leveling with this new method of correlation spectroscopy. TOC-729-1 is acting as a reference when placed on an optical table, and TOC-729-3 is placed on two locations successively with geometric height difference (relative to TOC-729-1) -0.792(10) m and 1.072(10) m, respectively. As shown in Fig.4(a) and Table 1, at two different elevations, the frequency differences between TOC-729-1 and TOC-729-3 are measured as -50.5(3.3) mHz and 39.1(2.6) mHz, respectively. Most of the systematic frequency shift evaluation process is the same as in our previous work\cite{YuanjinboAPL,liu2023laboratory}. Since only the information of the frequency difference of the two optical clocks can be obtained in the clocks' operation with correlation spectroscopy, frequency shifts caused by servo, electric quadrupole, and magnetic field drifts are evaluated in differential mode. After correcting the systemic shifts of -15.2(6.4) mHz and -12.8(7.7) mHz, the absolute frequency differences at two different elevations are synthesized to be -35.3(7.2) mHz and 51.9(8.2) mHz, respectively. The fractional frequency shifts are calculated as -8.6(1.8)$\times10^{-17}$ and -12.6(2.0)$\times10^{-17}$.

According to Einstein's general theory of relativity, the frequency change of the optical clock due to the change of geopotential difference can be expressed as $\Delta\nu/\nu_c = $g$\Delta h /c^2$ \cite{PhysRevLett.4.337} ,Where $\nu_c$ is the clock transition frequency,  $\Delta\nu$ is the clock frequency difference caused by the geopotential difference,  $g$ is the gravitational acceleration, $\Delta h$ is the elevation difference, and $c$ is the speed of light. According to the gravitational acceleration $g$=9.793461(2) m/s$^2$ at the locations of clocks, the elevation differences measured by the optical clocks are -78.8 (16.1) cm and 115.9 (18.3) cm, respectively. These results are consistent with the elevation differences obtained by the geometric measurement within the error range, and are also consistent with the results obtained by the traditional method of Rabi spectroscopy\cite{liu2023laboratory}.

\section{Conclusion}
As a conclusion, we demonstrate the closed-loop clock operation on the order of 100 ms using an interrogation laser with artificially destructed coherence time at the millisecond level based on the method of correlation spectroscopy. Frequency comparison with the instability of 6.0$\times10^{-15}$/$\sqrt{\tau/s}$ is achieved, which is 20 times better than that of the traditional Rabi spectroscopy using the same destructive laser. Our scheme is applied to an elevation difference measurements experiment using transportable optical clocks. This result has proved that correlation spectroscopy between independent optical clocks at different locations can improve the measurement precision beyond the laser coherence time significantly.

In this demonstration, the clock probe time is limited to 100 ms level due to the strong magnetic sensitivity of ${^{40}}$Ca${^+}$ ion. In the future, frequency comparison with correlation spectroscopy can be applied to other clock systems that are very insensitive to magnetic field noise, such as ${^{176}}$Lu${^+}$, ${^{171}}$Yb${^+}$(E3 transition), ${^{27}}$Al${^+}$, and of course neutral atoms like ${^{87}}$Sr, ${^{171}}$Yb, and so on\cite{RevModPhys.87.637}. Considering that the clock lifetime of these atoms is in tens of seconds or even longer, the clock probe time can be easily extended to several seconds or even tens of seconds using correlation spectroscopy, which means that the instability of the frequency comparison can reach the level of 1$\times10^{-16}$/$\sqrt{\tau/s}$ or better. This excellent stability of frequency comparison accessing a statistical uncertainty of 1$\times10^{-18}$ in several hours opens a new avenue for geodetic surveys, which resolves an elevation difference of one centimeter in a short averaging time. Correlation spectroscopy is suitable for the frequency difference measurement of optical clocks in complex environments outside the laboratory, which greatly reduces the dependence of optical clock network on ultra-stable lasers, and has a bright application prospect in the field of  chronometric leveling.

\begin{acknowledgments}
We thank Dr. Z. Zhang for helpful comments on the manuscript. This work was supported by the Basic Frontier Science Research Program of Chinese Academy of Sciences (Grant No. ZDBS-LY-DQC028), the National Key Research and Development Program of China (Grant No. 2017YFA0304404), and the National Natural Science Foundation of China (Grant No. 11674357 and U21A20431).
\end{acknowledgments}

\nocite{*}

\bibliography{apssamp_1}
\end{document}